\documentclass[twocolumn,tighten]{aastex63}
\pdfoutput=1 %for arXiv submission
\usepackage{amsmath,amstext}
\usepackage{apjfonts}
\usepackage{afterpage}
\usepackage{xcolor}
\usepackage{threeparttable}
\usepackage{graphicx}
\usepackage{epsfig}
\usepackage{booktabs}
\usepackage{multirow}
\usepackage{float}
\usepackage{mathrsfs}
\DeclareUnicodeCharacter{1E11}{\k{a}}
\begin{document}
\author[0009-0003-6348-7143]{Tong Zhao}
\affiliation{School of Physics, Peking University, Beijing 100871, China}

\author[0000-0002-2749-6638]{Mingyu Ge}
\affiliation{State Key Laboratory of Particle Astrophysics, Institute of High Energy Physics, Chinese Academy of Sciences, Beijing 100049, China}

\author[0000-0002-9042-3044]{Renxin Xu}
\affiliation{School of Physics, Peking University, Beijing 100871, China}

\title{An Intrinsic Degeneracy in a Simplified Semi-Analytic Two-Spot Model for Thermal X-Ray Pulse Profiles}

\begin{abstract}
We identify an intrinsic degeneracy in the simplified semi-analytic two-spot model for parameter inference in thermal X-ray pulse profile modeling. Although exact algebraic degeneracy exists in our simplified model without Doppler effects, we generate synthetic pulse profiles and show that this degeneracy can still result in two peaks on the likelihood surface if Doppler effects, statistic errors, and the instrument response are considered. The analytic degeneracy we discovered may provide intuition for multi-modal posterior distributions discovered in previous work~\citep{Vinciguerra2023,Vinciguerra2024,Miller2019,Miller2021}. However, whether the effect of this degeneracy persists in real observational posteriors with background models and more complex spot geometries still requires further systematic study.
\end{abstract}


\section{Introduction}
Measurements of neutron star masses and radii provide crucial constraints on the equation of state of cold dense matter. Among the various techniques, pulse profile modeling of rotation-powered millisecond pulsars has emerged as a powerful method, particularly with observations from the Neutron Star Interior Composition Explorer (NICER;~\citealt{NICERins}) and future observations from the enhanced X-ray Timing and Polarimetry (eXTP) mission, which is scheduled to launch in early 2030~\citep{eXTP2025}.

The standard approach in these analyses employs numerical ray-tracing to compute the observed flux from hot regions on the neutron star surface, often assuming two circular hot spots with independent parameters. However, numerical methods cannot interpret the origin of parameter degeneracies and also make it difficult for Bayesian inference to exhaustively explore the parameter space because of the large computational expense. In studies based on synthetic pulse profiles, multi-modal structures in the posterior surface have already been discovered by~\citet{Vinciguerra2023,Vinciguerra2024,Miller2019,Miller2021}. However, why such modes arise is not interpreted. In previous work, we developed a semi-analytic model with two antipodal hot spots under the Schwarzschild plus Doppler (S+D) approximation~\citep{Pout1,Zhao2024,Zhao2025}, which allows us to study the impact of degeneracies and wrong assumptions on parameter inference.

In this paper, we extend that semi-analytic framework to the more general case with two non-antipodal hot spots, making it resemble the numerical ST-U models widely used in pulse profile modeling. Assuming the spots are sufficiently small so that each can be treated as a point source, we derive analytic expressions for the observed flux that depend on the geometric parameters through a set of dimensionless parameter combinations. This formulation reveals an intrinsic degeneracy: two distinct sets of parameter values ($u$, $\theta$, $\zeta_1$, $\zeta_2$) can produce identical pulse profiles. We generate synthetic pulse profiles and show that even if Doppler effects and the instrument response are considered, this degeneracy still leads to another peak on the posterior surface close to the other solution in the parameter inference.

This paper is organized as follows: In Section~\ref{sec:geometry} we present our generalized semi-analytic two-spot model and define the relevant parameters in the model. In Section~\ref{sec:degeneracy} we derive the analytic degeneracy and show how it arises from the structure of the flux equation. In Section~\ref{sec:multimodal} we demonstrate that this degeneracy results in multiple peaks on the posterior surface if there is insufficient prior knowledge for $u$, $\theta$, $\zeta_1$, and $\zeta_2$ by fitting our model to synthetic pulse profiles. Finally, we discuss the implications for existing NICER results. We conclude with a summary in Section~\ref{sec:summary}.

\section{Geometry and Parameterization}
\label{sec:geometry}
\begin{figure}[t]
	\centering
	\includegraphics[width=1\linewidth, keepaspectratio]{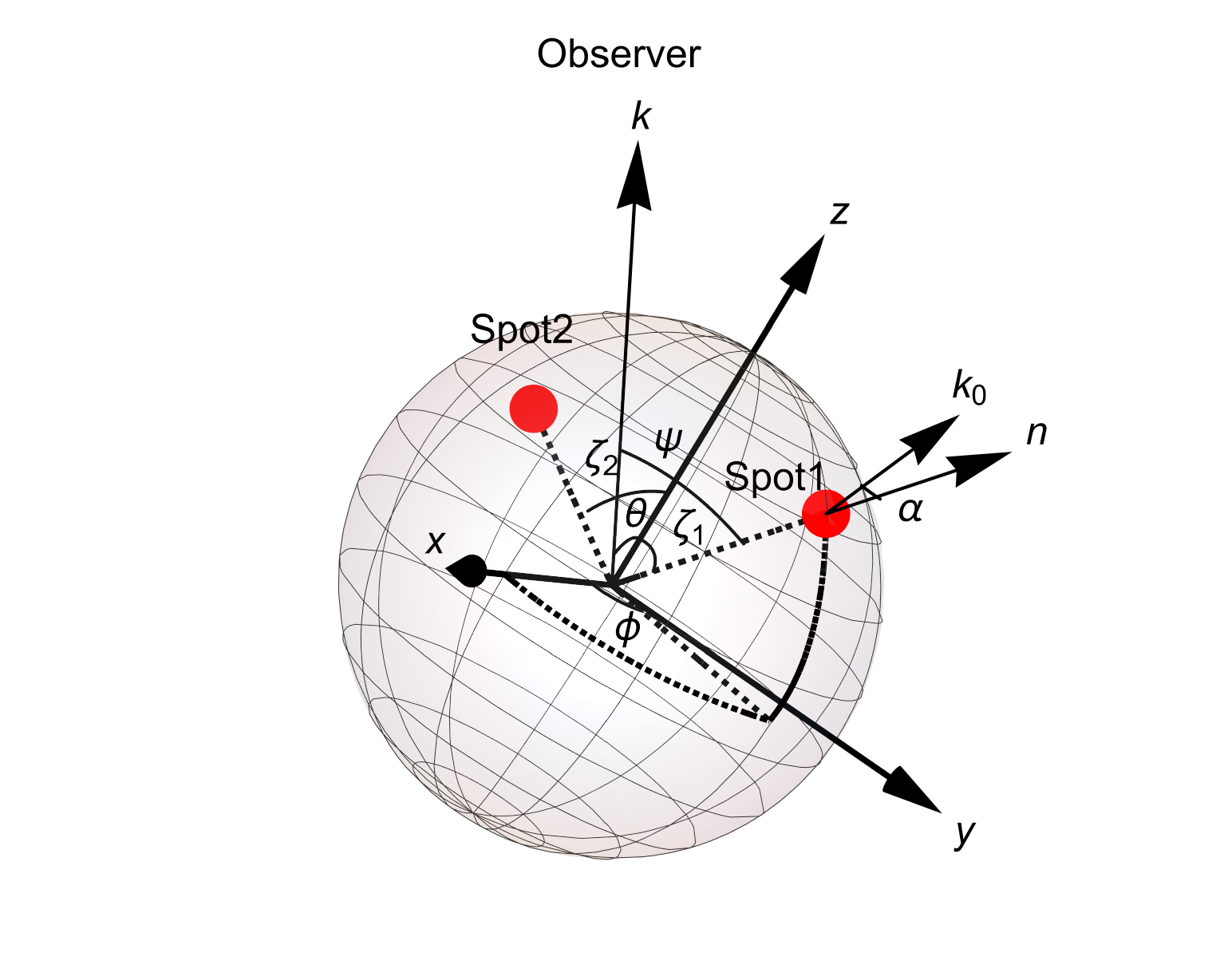}
	\caption{The geometry sketch for ST-U and our semi-analytic two-spot model. $\hat{k}$ is the unit vector pointing from the center of the neutron star to the observer. $\hat{k_0}$ is the moment vector of the photon emitted from the hot spot. $\hat{n}$ is the surface normal at the center of the hot spot. $\alpha$ is the emission angle, the angle between $\hat{k_0}$ and $\hat{n}$. $\theta$ is the observer inclination angle, the angle between $\hat{k}$ and the z-axis. The z-axis is the assumed spin axis of the neutron star. $\psi$ is the inflection angle, the angle between $\hat{n}$ and $\hat{k}$. $\phi$ is the neutron star phase angle. $\zeta_1$ and $\zeta_2$ are the colatitude angles of the center of each hot spot, respectively.
    }
    \label{fig:geometry}
\end{figure}

A widely used configuration of the emission region in pulse profile models assumes two single-temperature, uniform-temperature emission regions with independent parameters. That is, two circular hot regions on the neutron star surface, each with different inclination angles, phase angles, effective temperatures, and opening angles. However, within each hot region, the temperature distribution is assumed to be uniform. In NICER observations, models with this kind of emission region are often referred to as ST-U models~\citep{Riley2019,Miller2019,Riley2021,Miller2021,Choudhury2024,Salmi2024,Mauviard2025}. Different groups may use different terminology for related two-spot models, while the ST-U model in this work refers to the model we described above. The geometry sketch for ST-U and our semi-analytic model is shown in Fig.~\ref{fig:geometry}. $\hat{k}$ is the unit vector that points from the center of the neutron star to the observer. $\hat{k_0}$ is the moment vector of the photon emitted from the hot spot. $\hat{n}$ is the surface normal in the center of the hot spot. $\alpha$ is the emission angle, the angle between $\hat{k_0}$ and $\hat{n}$. $\theta$ is the angle of inclination of the observer, the angle between $\hat{k}$ and the z-axis. $\psi$ is the bending angle, the angle between $\hat{n}$ and $\hat{k}$, indicating how much the light is bent toward the observer. $\phi$ is the neutron star phase angle. $\zeta_1$ and $\zeta_2$ are the colatitude angles of the center of each hot spot, respectively.

In this paper, we generalize our semi-analytic antipodal two-spot model in~\citet{Zhao2024} to the non-antipodal case to resemble the numerical ST-U model. We further assume that both spots are small enough, and thus the emission can be considered to originate from the center of the spot. This is a good approximation because the error is less than $1\%$ if the opening angle of the spot is $\lesssim 5^\circ$~\citep{Hajime2019}. For convenience, we assume that the observer is in the sky of the north hemisphere and that the initial phase angle of the primary spot is zero. Thus, there are twelve parameters in the entire model, listed in Table~\ref{table:1}.

\begin{table}[t]
\caption{Physical Parameters for Neutron-Star Pulse Profile Modeling}
\begin{center}
%\begin{threeparttable}
\begin{tabular}{c c c c}
\toprule
Parameter & Description & Units & Range \\
\midrule
$M$ & Neutron star mass & $M_\odot$ & $\sim 1-2.5$ \\ 
$R$ & Neutron star radius & km & $\sim 5-20$ \\ 
$\theta$ & Observer inclination angle & degrees & $0-90$ \\
$\zeta_1$ & Colatitude of the first spot& degrees & $0-180$ \\
$\zeta_2$ & Colatitude of the second spot& degrees & $0-180$ \\
$\phi_2$ & Phase angle of the second spot& degrees & $0-360$ \\
$T_1$ & Effective temperature & keV & $>0$\\
$T_2$ & Effective temperature & keV & $>0$\\
$dS_1$ & Spot area of the first spot & m$^2$ & $>0$ \\ 
$dS_2$ & Spot area of the second spot & m$^2$ & $>0$ \\ 
$D$ & Distance & kpc & $>0$ \\ 
$N_H$ & Neutral H column density & $10^{20}$~cm$^{-2}$ & $>0$ \\ 
\bottomrule
\end{tabular}
%\end{threeparttable}
\end{center}
\label{table:1}
\end{table}

To reduce the correlation between parameters we need to parameterize the model carefully and notice that: First, $M$ and $R$ are usually not well-constrained independently given the low spin frequency of neutron stars and the data quality of NICER observations. Under the S+D approximation, where the space time depends only on compactness $u=2GM/Rc^2$, measuring the strength of Doppler effects is the only way to constrain the radius. Thus, meaningful constraints on both mass and radius are usually obtained for neutron stars with frequencies $\gtrsim300$~Hz and with large inclination angle and spot colatitude~\citep{Lo2013,MRdegeneracy2014,Zhao2024}. Therefore, prior knowledge of $M$ is usually included in the parameter inference~\citep{Riley2021, Miller2021,Choudhury2024,Salmi2024}. Second, the total area of all hot spots or the emission region $dS^\prime$ and the distance $D$ cannot be independently restricted because they contribute only an overall factor $dS^\prime/D^2$, the solid angle, to the total flux. Here, we use primed symbols to denote the quantities measured on the neutron star surface. This can be well understood following the idea of analytic models. The hot spots or the emission region can be divided into small segments, and the total flux is the sum of all fluxes from these small segments:~$F_{total}=\sum_i F_i$. For each small segment, the assumption of the single-spot analytic model is valid. Thus, $F_i=dS^\prime_i/D^2*f_i$ is proportional to $dS^\prime_i/D^2$. Then, we can define the total area $dS^\prime=\sum_idS^\prime_i$, and $F_{total}=\sum_i dS^\prime_i/D^2*f_i=dS/D^2*\sum_ir_if_i$, where $r_i=dS^\prime_i/dS^\prime$. We can further extract $r_1$ from the equation, define $a_i=r_i/r_1$ and get $F_{total}=dS^\prime_1/D^2*(1+\sum_{i>1}a_if_i)$. If there are $n$ small segments, eventually, the $n+1$ free parameters $dS^\prime_1$, $dS^\prime_2$... $dS^\prime_n$ and $D$ become $n$ free parameters $a_2$, $a_3$...$a_n$ and $dS^\prime_1/D^2$. 

Therefore, in this work, we fix the mass $M$ and define $A=dS^\prime_1/D^2$ and $a_r=dS^\prime_2/dS^\prime_1$. For convenience, we also define $\Delta \phi$ as the longitudinal phase offset relative to antipodal longitude: $\phi_2=180^\circ+\Delta \phi$ rather than directly using $\phi_2$. We also rescale $A$ by a fiducial value $A_0$ to avoid extremely small values. Finally, we define some dimensionless axillary parameters to simplify the equations. They are the compactness $u=2GM/Rc^2$, $q_1$, $q_2$, $p_1$ and $p_2$ relative to gravitational self-lensing:
\begin{equation}
q_1=u+(1-u)\cos\theta \cos\zeta_1,
\label{eq:q1}
\end{equation}
\begin{equation}
q_2=u+(1-u)\cos\theta \cos\zeta_2,
\label{eq:q2}
\end{equation}
\begin{equation}
p_1=(1-u)\sin\theta \sin\zeta_1,
\label{eq:p1}
\end{equation}
\begin{equation}
p_2=(1-u)\sin\theta \sin\zeta_2.
\label{eq:p2}
\end{equation}

Furthermore, we fixed the column density of neutral hydrogen $N_H$ to reduce irrelevant degrees of freedom as much as possible. Thus, the original twelve parameters can be reduced to nine free parameters listed in Table~\ref{table:2}. The compactness is assumed to be less than 0.5 because the approximate light bending equation is valid only if $u<0.5$. Also, neutron stars with $u>0.5$ are not observed in the literature, and large compactness will increase the second-order gravitational lens effect that is neglected in both our model and the ST-U model.
\begin{table*}[t]
\caption{Free Parameters for Flux of our Semi-analytic Two-spot Model}
\begin{center}
%\begin{threeparttable}
\begin{tabular}{c c c c}
\toprule
Parameter & Description & Units & Range \\
\midrule
$u$ & Compactness & dimensionless & $\sim 0.2-0.5$ \\ 
$\theta$ & Observer inclination angle & Degrees & $0-90$ \\
$\zeta_1$ & Colatitude of the first spot& Degrees & $0-180$ \\
$\zeta_2$ & Colatitude of the second spot& Degrees & $0-180$ \\
$\Delta\phi$ & Second spot phase Angle deviation& Degrees & $-180-+180$ \\
$T_1$ & Effective temperature & keV & $>0$\\
$T_2$ & Effective temperature & keV & $>0$\\
$a_r$ & Area ratio of the second spot to the first Spot & Dimensionless & $>0$ \\ 
$A$ & $\frac{dS_1}{D^2}/A_0$ & Dimensionless & $>0$ \\ 
\\ 
\bottomrule
\end{tabular}
%\end{threeparttable}
\end{center}
\label{table:2}
\end{table*}

\section{The Intrinsic Degeneracy in the Semi-analytic Two-spot Model}
\label{sec:degeneracy}
In this section, we present a simplified model to explain why there is a degeneracy in the simplified two-spot model. In the next section, we will show that this degeneracy may lead to multi-modal structures in the posterior surface during parameter inference of synthetic and real observational data. Under the S+D approximation and ignoring the Doppler effect if the neutron star frequency is $f\lesssim200$~Hz, the flux from the first spot is given by~\citep{Zhao2024}:
\begin{equation}
F_1(E,\phi)=
\begin{cases}
    {\cal F}_1(E,\phi)\;, &\text{if~}~q_1+p_1\cos\phi>0\\
    0\;, & \text{otherwise}
    \label{eq:F_1}
\end{cases}
\end{equation}
where
\begin{equation}
\begin{split}
&{\cal F}_1(E,\phi)=q_1(1+\bar{h}_1q_1)\\
&\times\bar{F}_1(E)\left[1+r_1(E)\cos\phi+r_2(E)\cos\phi^2\right]
\label{eq:Fourier}
\end{split}
\end{equation}
\begin{equation}
\bar{F}_1(E)=\frac{A(1-u)^{3/2}}{1+(2/3)\bar{h}_1} I^\prime _b,
\label{eq:Fbar}
\end{equation}
\begin{equation}
r_1(E)=\frac{p_1}{q_1}\left(1+\frac{\bar{h}_1q_1}{1+\bar{h}_1q_1}\right),
\label{eq:r1}
\end{equation}
\begin{equation}
r_2(E)=\left(\frac{p_1}{q_1}\right)^2\frac{\bar{h}_1q_1}{1+\bar{h}_1q_1},
\label{eq:r2}
\end{equation}
and the beaming factor $\bar{h}$ is evaluated at
\begin{equation}
    \bar{h}_1=\bar{h}(E/\sqrt{1-u},T_1)=\bar{h}(\frac{E}{T_1\sqrt{1-u}}).
\end{equation}

$I^\prime_b$ is the blackbody emission intensity from the neutron star surface:
\begin{equation}
I^\prime_b(E,T)=\frac{2\frac{E^3}{\sqrt{1-u}^3}}{h^3c^2(e^{E/kT_1\sqrt{1-u}}-1)}.
\end{equation}
Here, we want to clarify that our beaming pattern is not a simple blackbody emission. In general, the specific intensity of the emission $I^\prime(E^\prime,T,\alpha)$ is a function of the energy of the emitted photon, the local effective temperature, and the emission angle. It depends on atmospheric models, and there is no analytic formula for it. In the semi-analytic model, we adopt a parameterized function of $E$, $\alpha$ and $T$ to describe it. First, we make a Taylor expansion of $I^\prime/I_b$ to the first order of $\cos\alpha$ so that $I^\prime\approx(1+\bar{h}\cos\alpha)I_b$. Then, normalize the total flux according to the total flux of the blackbody emission. We have
\begin{equation}
I^\prime\approx\frac{(1+\bar{h}\cos\alpha)I_b}{1+2/3\bar{h}},
\end{equation}
\begin{equation}
h(E^\prime,T)=a+b\left(\frac{E^\prime}{kT}\right)+c\left(\frac{E^\prime}{kT}\right)^2\;.
\label{eq:beaming_energy}
\end{equation}
Here, $a$, $b$ and $c$ are obtained by fitting this formula to a real atmosphere model~\citep{atmosphere}. Certainly, we can keep more orders of $\cos\alpha$, but the numerical result shows that keeping the first order is already a good approximation~\citep{Zhao2024}. Hence, the blackbody emission intensity $I_b$ here is used as a normalization factor for the intensity not based on physical reasons. And our effective temperature is defined as the equivalent temperature if we equate the total emission to the total emission of a blackbody. Thus, the emission pattern in our model is an approximation of the real atmosphere model in~\citet{atmosphere}, but the effective temperature is the equivalent blackbody temperature that is different from the temperature in ionized hydrogen atmosphere models.

Notice that the factor $(1-u)^{3/2}$ in the numerator of equation~\ref{eq:Fbar} cancels perfectly with the factor $(1-u)^{3/2}$ in $I^\prime_b$. In addition, we can define $T_{1\infty}=T_1\sqrt{1-u}$ to incorporate $T_1$ and $\sqrt{1-u}$ into a new parameter $T_{1\infty}$. Therefore, we can simplify~\ref{eq:Fbar} to
\begin{equation}
\bar{F}_1(E)=\frac{2AE^3}{h^3c^2(1+\frac{2}{3}\bar{h}_1)(e^{\frac{E}{kT_{1\infty}}}-1)}.
\end{equation}
Similarly, the flux from the second spot is given by:
\begin{equation}
F_2(E,\phi)=
\begin{cases}
    {\cal F}_2(E,\phi)\;, &\text{if~}~q_2-p_2\cos(\phi+\Delta\phi)>0\\
    0\;, & \text{otherwise}
    \label{eq:F_2}
\end{cases}
\end{equation}
where
\begin{equation}
\begin{split}
&{\cal F}_2(E,\phi)=q_2(1+\bar{h}_2q_2)\\
&\bar{F}_2(E)\left[1-s_1(E)\cos(\phi+\Delta\phi)+s_2(E)\cos(\phi+\Delta\phi)^2\right]
\label{eq:Fbar2}
\end{split}
\end{equation}
\begin{equation}
\bar{F}_2(E)=\frac{2a_rAE^3}{h^3c^2(1+\frac{2}{3}\bar{h}_2)(e^{\frac{E}{kT_{2\infty}}}-1)},
\end{equation}
\begin{equation}
s_1(E)=\frac{p_2}{q_2}\left(1+\frac{\bar{h}_2q_2}{1+\bar{h}_2q_2}\right)\;,
\label{eq:s1}
\end{equation}
\begin{equation}
s_2(E)=\left(\frac{p_2}{q_2}\right)^2\frac{\bar{h}_2q_2}{1+\bar{h}q_2}\;,
\label{eq:s2}
\end{equation}
and the beaming factor $h$ is evaluated at
\begin{equation}
    \bar{h}_2=\bar{h}(\frac{E}{T_{2\infty}}).
\end{equation}
The total flux is the sum of the two fluxes:
\begin{equation}
F(E,\phi)=F_1(E,\phi)+F_2(E,\phi). 
\end{equation}

Here, we can see that $u$, $\theta$, $\zeta_1$ and $\zeta_2$ are completely absorbed into the four dimensionless parameters $p_1$, $p_2$, $q_1$, and $q_2$. In addition, the total flux can be expressed by the nine new parameters: $q_1$, $q_2$, $p_1$, $p_2$, $\Delta\phi$, $A$, $a_r$, $T_{1\infty}$, and $T_{2\infty}$. Hence, if a different set of parameter values gives us the same $q_1$, $q_2$, $p_1$, $p_2$, $\Delta\phi$, $A$, $a_r$, $T_{1\infty}$, and $T_{2\infty}$, even if $u$, $\theta$, $\zeta_1$, and $\zeta_2$ are completely different, the total flux, and therefore the final pulse profile, will be exactly the same. By solving equations~\ref{eq:q1},~\ref{eq:q2},~\ref{eq:p1} and~\ref{eq:p2} for $u$, $\theta$, $\zeta_1$, and $\zeta_2$, we can show that this is possible. This set of equations usually has two solutions, which means that two configurations or modes can result in the same flux.

Although analytical solutions for $\theta$, $\zeta_1$ and $\zeta_2$ are difficult to express, by combining equations~\ref{eq:q1},~\ref{eq:q2},~\ref{eq:p1}, and~\ref{eq:p2}, the compactness $u$ can be implicitly expressed by the solution of a quadratic equation:
\begin{equation}
C_2u^2+C_1u+C_0=0,
\label{eq:quadratic}
\end{equation}
where
\begin{equation}
\begin{split}
&C_2=K^2-4\Delta(K-P),\\
&C_1=-2P(\Delta\Sigma+K)-2\Delta Q+2\Delta K(\Sigma+1),\\
&C_0=(\Delta\Sigma+K)Q-\Delta\Sigma K,\\
\end{split}
\end{equation}
and
\begin{equation}
\begin{split}
&K=p_2^2-p_1^2,\\
&P=p_2^2q_1-p_1^2q_2,\\
&Q=p_2^2q_1^2-p_1^2q_2^2,\\
&\Delta=q_1-q_2,\\
&\Sigma=q_1+q_2.
\end{split}
\end{equation}
Therefore, when equation~\ref{eq:quadratic} has two physical solutions, there are two degenerate sets of parameter values that give the same flux. This degeneracy will result in two peaks on the likelihood surface if there is no prior knowledge to break it. However, if the two solutions are too close to each other, or if one solution for $u$ is not reasonable, this degeneracy does not matter.

For the two-antipodal-spot model, we have $\zeta_2=180^\circ-\zeta_1$. There is a degeneracy related to the interchange between the observer inclination angle $\theta$ and the magnetic inclination angle $\zeta_1$. This degeneracy can be considered as the special case of the degeneracy we find in the non-antipodal model. Consider equations~\ref{eq:q1},~\ref{eq:q2},~\ref{eq:p1} and~\ref{eq:p2}. If the two spots are antipodal, $\Delta\phi=0$, $p_2=p_1$ and $q_2=-q_1$. The problem then reduces to looking for the other solution for $u$, $\theta$, and $\zeta_1$ to get the same $q_1$ and $p_1$. Obviously, keeping the same $u$ and swapping $\theta$ and $\zeta_1$ is this solution. In addition, we can calculate the discriminant of~\ref{eq:quadratic}: $4(p_1^2-p_2^2)^2 (p_1^2-(-1+q_1)^2)(p_2^2-(-1+q_2)^2)(q_1-q_2)^2$. It is zero when $p_2=p_1$, which means that the two solutions share the same $u$ in the antipodal-spot model.

\section{Two High Likelihood Regions on the Likelihood Surface}
\label{sec:multimodal}

In the last section, the exact algebraic degeneracy was derived under assumptions such as negligible Doppler effects and point-like spots. However, in real observations, Doppler effects cannot be ignored, especially for neutron stars with frequency $f\gtrsim 300$~Hz. In this section, we will make synthetic pulse profiles and show that this degeneracy can still result in two high likelihood regions corresponding to two modes when Doppler effects and instrument response are taken into account. \textbf{We define the mode that corresponds to the ground truth as the primary mode, the other mode closed to the other solution in the simplified model as the secondary mode}. Both modes can fit the synthetic pulse profile well, but the primary mode is statistically more important than the secondary mode according to our estimation of the posterior mass ratio. In the Appendix, we also show how the result changes with the frequency and the total number of photons of the data. However, whether these two modes can be distinguished in real observation with background models requires further study.

For the synthetic data, we calculate the observed flux with our semi-analytic two-spot model. To mimic real observational data, we generate multi-energy synthetic pulse profiles and consider Doppler effects, the interstellar extinction, and the effective area and response matrix of NICER to get the final observed photon counts. The interstellar extinction parameters and the instrument response are the same as in our previous work~\citep{Zhao2024}. The photon energy range, the exposure time, and the total number of photons are the same as those used by~\citet{Miller2021}. The range of photon energy is between 0.3~keV and 1.5~keV. For our synthetic data, energy channels are binned to ensure a sufficient number of photons per channel, and Gaussian noise is added to each channel. The width of our energy channel is 0.05~keV, which is ten times the width of the energy channels for the NICER data. This will significantly increase computational efficiency. For each energy channel, the pulse profile has distinctive features because enough photon counts are included, while for the original energy channel, most single-channel pulse profiles are overwhelmed by noise. Pulse profiles are within one spin period that is divided into 32 phase bins. Because strict priors for the mass value are usually known in parameter inference for radius measurement, we fixed the mass $M$ at two solar masses and adopted the radius $R$ instead of $u$ as a free parameter. We define $A_0=dS/D^2$ where $dS=1.04\times 10^6$~m$^2$ and $D=1.099$~kpc as a fiducial value $A_0$. In our model, we adopt two parameters for area and distance: the area of the first hot spot divided by distance squared, which is rescaled by this fiducial value $A=dS_1/D^2/A_0$, and the ratio of the second hot spot area to the first hot spot area $a_r=dS_2/dS_1$. For the effective temperature, because our emission is normalized to the total flux of a blackbody emission, the effective temperature is larger than that used in models with an ionized hydrogen atmosphere. Thus, to make our synthetic pulse profiles, we assume $T_1=0.15$~keV and $T_2=0.14$~keV. Finally, the column density of neutral hydrogen is also fixed at $0.226 \times10^{20}$~cm$^{-2}$. All free parameters and their injected values for synthetic pulse profiles are listed in Table~\ref{table:3}.

In this section, we consider Doppler effects and assume that the frequency is 300~Hz. The Doppler factor $\delta_D$ gives several corrections to the flux equation. First, the photon energy emitted from the neutron star surface is $E/\delta_D\sqrt{1-u}$. Second, the values of emission angle and spot area measured on the surface of the neutron star are different from the observed value. However, if the frequency is about 300~Hz, all these corrections and the time delay effect are very small because they involve only $\delta_D$ to the first power, which is very close to 1. What really matters is the overall factor $\delta_D^4$, a fourth order term arises from the Doppler effect in the flux equation~\citep{Zhao2024}. Thus, we introduce the leading term of this factor as the Doppler effect correction:
\begin{equation}
\delta_D^4\approx 1-\frac{8\pi \nu R\sin\zeta\sin\theta\sin\phi}{c},
\end{equation}
where $\nu$ is the neutron star frequency and $c$ is the speed of light. $\zeta$ and $\phi$ are the colatitude and phase angle of the hot spot, so they take different values for the two different hot spots.

\begin{table*}[t]
\caption{Physical Free Parameters in Our Semi-analytic Two-spot Model}
\begin{center}
%\begin{threeparttable}
\begin{tabular}{c c c c}
\toprule
Parameter & Description & Units & assumed value \\
\midrule
$R$ & Neutron star radius & km & $13.109$~km \\ 
$\theta$ & Observer inclination angle & degrees & $70$ \\
$\zeta_1$ & Colatitude of the first spot& degrees & $115$ \\
$\zeta_2$ & Colatitude of the second spot& degrees & $140$ \\
$\Delta\phi$ & Phase angle deviation of the second spot& degrees & $-12$ \\
$T_1$ & Effective temperature & keV & $0.15$\\
$T_2$ & Effective temperature & keV & $0.14$\\
$a_r$ & Area ratio of the second spot to the first & dimensionless & $1.5$ \\ 
$A$ & $dS_1/D^2/A_0$ & dimensionless & $1.756$ \\ 
\bottomrule
\end{tabular}
%\end{threeparttable}
\end{center}
\label{table:3}
\end{table*}

\begin{figure*}
	\centering
	\includegraphics[width=1\linewidth, keepaspectratio]{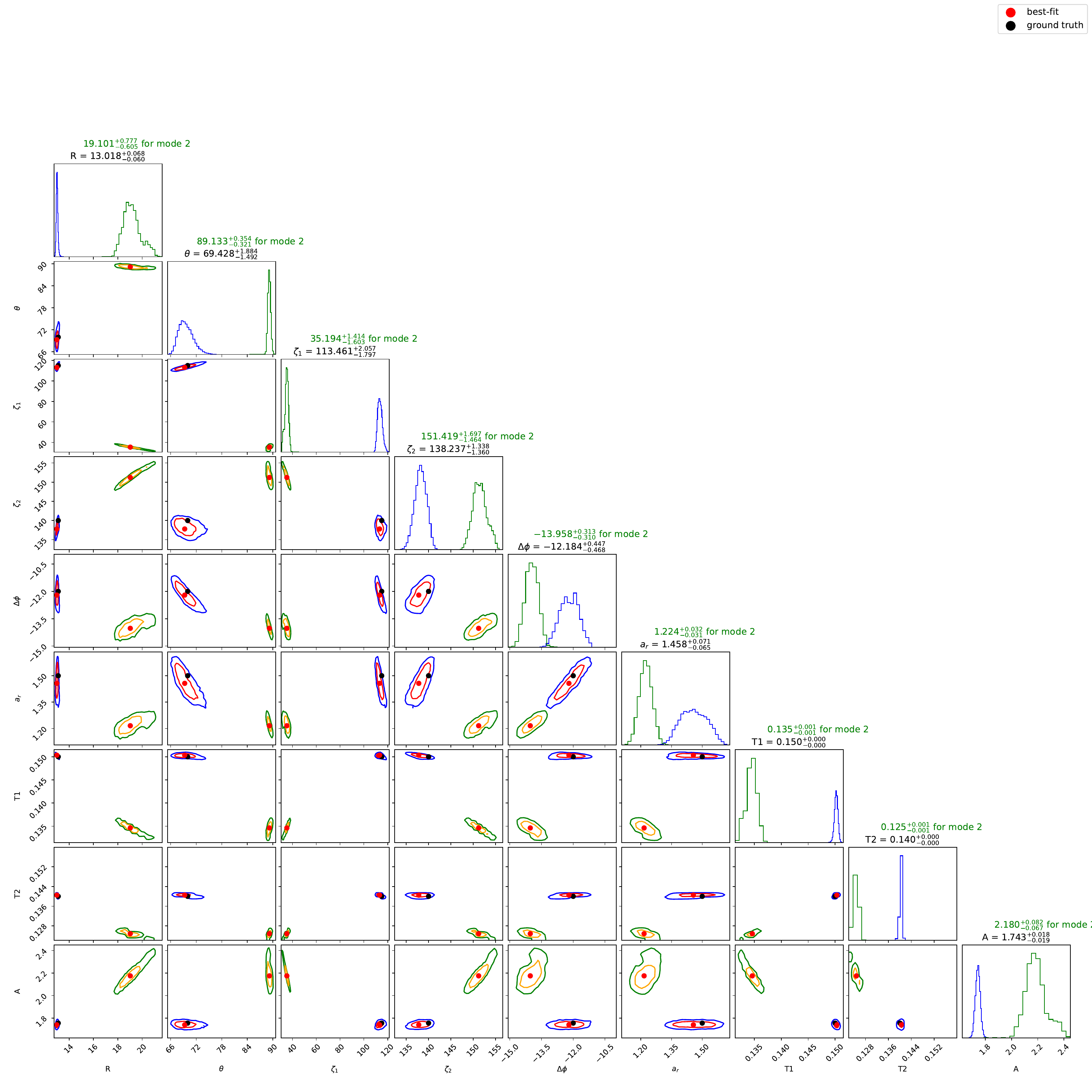}
	\caption{The corner plot of two results with the same model fitting to the same synthetic data. We run the MCMC twice starting from two initial values to sample and end up with two peaks on the likelihood surface. The first best-fit point is close to our assumed value but the second one is completely different. 68-th and 95-th percentile contours for the first run are shown in red and blue, and for the second run in orange and green. 
    }
    \label{fig:corner}
\end{figure*}

\begin{figure*}
	\centering
	\includegraphics[width=1\linewidth, keepaspectratio]{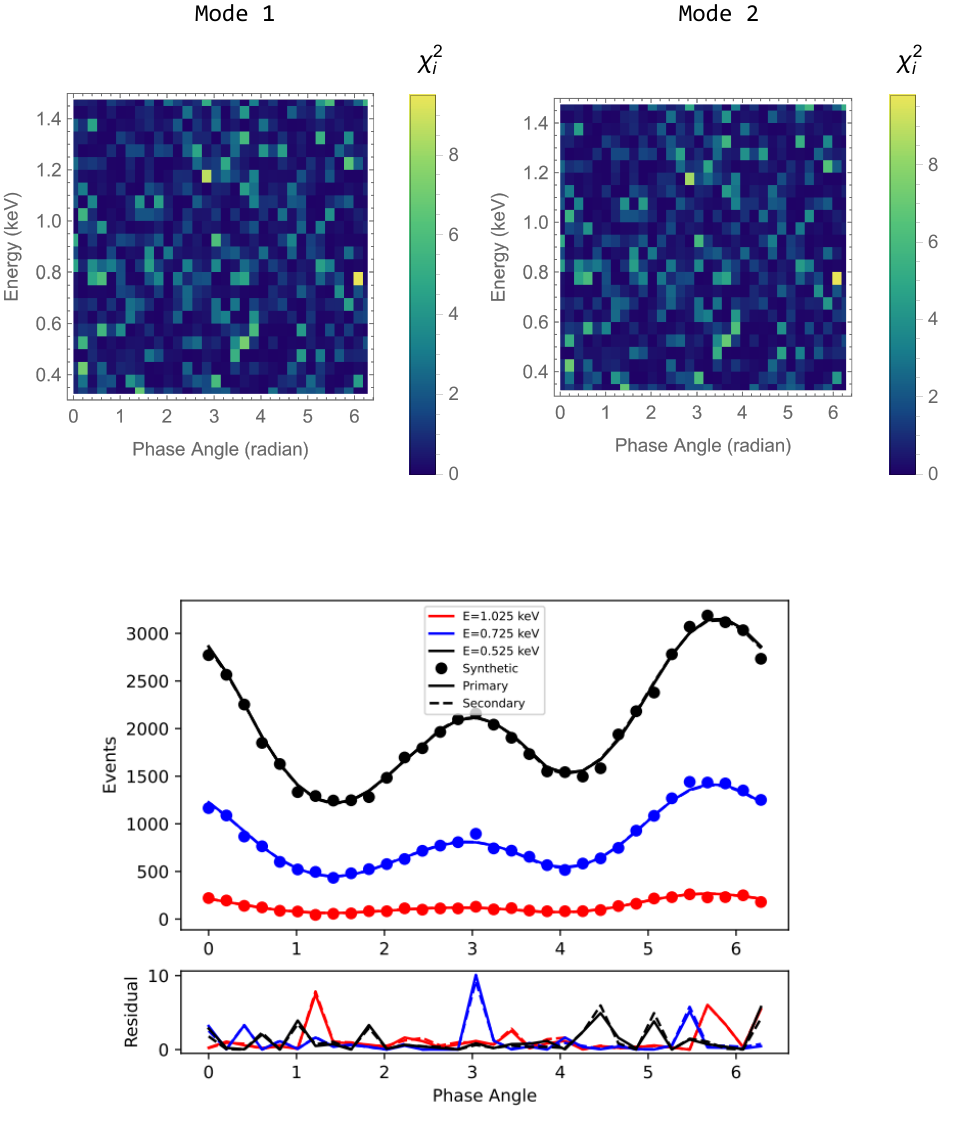}
	\caption{The comparison of the two-dimensional data corresponding to the two modes with the synthetic data. The difference between the model and the synthetic data in all energy and phase bins. Each pixel is colored according to the residual defined as $\chi_i^2=(D_i-M_i)^2/\sigma_i^2$, where $D_i$ is the number of photons given by the synthetic data, $M_i$ is the number of photons given by the model, and $\sigma_i$ is the error of data. The top left panel is for the primary mode that is close to the ground truth and the top right panel is for the secondary mode. We also show a few pulse profiles for $E=1.025$~keV, $E=0.725$, and $E=0.525$~keV in the bottom panel of the same plot.
    }
    \label{fig:pulse-profiles}
\end{figure*}

We use a flat prior and a Gaussian likelihood function and start Markov-Chain Monte Carlo (MCMC) sampling from two different initial values for $u$, $\theta$, $\zeta_1$, and $\zeta_2$ close to the two solutions that lead to the same flux to find the two modes on the likelihood surface. A random walker is used to sample the parameter space with a step size of about $0.05\%$ of the initial parameter value. We check the convergence of the MCMC result by adding more chains until the result remains unchanged, and finally run ten million chains. The 68-th and 95-th percentile contours for the two runs are shown together in the same corner plot in Fig.~\ref{fig:corner}. The two runs converge to two best-fit points corresponding to the two modes where the likelihood function reaches local maxima. We also make a two-dimensional comparison of the residuals for the two modes in Fig.~\ref{fig:pulse-profiles}. We use a $\chi^2$-like residual $\chi_i^2=(D_i-M_i)^2/\sigma_i^2$, where $D_i$ is the number of photons given by the synthetic data, $M_i$ is the number of photons given by the model and $\sigma_i$ is the error of the data. In addition, we show a few typical pulse profiles illustrating the difference between the two modes in the bottom panel of the same plot.

\begin{figure*}[t]
	\centering
	\includegraphics[width=1\linewidth, keepaspectratio]{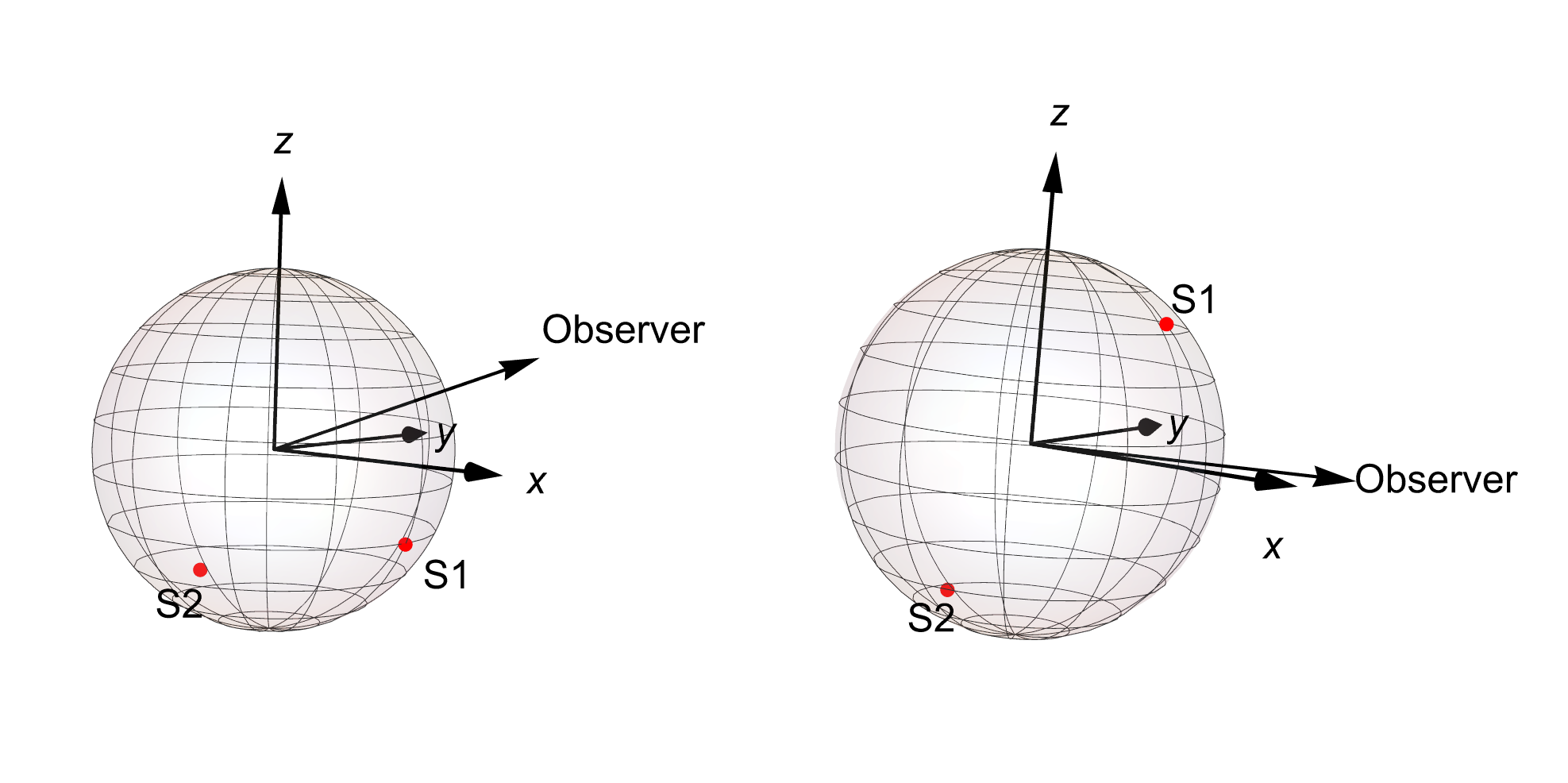}
	\caption{Geometry configurations of the two modes. The z-axis is the assumed spin axis of the neutron star. Position of two spots are presented by two red points, labeled as $S1$ and $S2$, and a vector is pointing from the neutron star center to the observer illustrating the direction of the observer.
    }
    \label{fig:configurations}
\end{figure*}

According to our discussion in the last section, the total flux is mainly determined by $q_1$, $q_2$, $p_1$ and $p_2$, and there are usually two sets of solutions for $u$, $\theta$, $\zeta_1$ and $\zeta_2$. The assumed values are $u=0.45$, $\theta=70^\circ$, $\zeta_1=115^\circ$, $\zeta_2=140^\circ$, and the corresponding radius is $R=13.109$~km. According to equation~\ref{eq:quadratic}, the compactness value of the other solution is $u=0.341$, which corresponds to $R=17.300$~km. The values for other parameters can be calculated numerically and we get $\theta=86.404^\circ$, $\zeta_1=45.460^\circ$, and $\zeta_2=149.634^\circ$. They correspond to the second best-fit in Fig.~\ref{fig:corner}, $R=19.101$~km, $\theta=89.133^\circ$, $\zeta_1=35.194^\circ$ and $\zeta_2=151.419^\circ$. In addition, the effective temperatures of the two best-fit points correspond to the same $T_\infty$ we defined. We also illustrate the geometry configurations of the two modes in Fig.~\ref{fig:configurations}.

We can see that if prior knowledge of $u$, $\theta$, or spot colatitude $\zeta_1$ and $\zeta_2$ is not provided, the effect of this degeneracy still exists even if Doppler effects and instrument response are considered. The degeneracy leads to two local high-likelihood regions, and the position of the second peak is close to the degenerate solution predicted by our simplified model without Doppler effects. With data ${\cal D}$, for the model ${\cal M}$ and parameters $\Theta$, our likelihood function is
\begin{equation}
{\cal L}({\cal D}|{\cal M},\Theta)=\exp(-\frac{1}{2}\chi^2)=\exp{[-\frac{1}{2}\sum_i\frac{({\cal M}_i-{\cal D}_i)^2}{\sigma_i^2}]}.
\end{equation}
The $\chi^2$ for the primary and secondary modes are 734.3235 and 739.2853, which correspond to the logarithmic likelihood values -367.1617 and -369.6426. Because we use the flat prior for all parameters, the logarithmic posterior satisfies
\begin{equation}
\ln({\cal P}({\cal M},\Theta|{\cal D}))=\ln({\cal L}({\cal D}|{\cal M},\Theta))+constant.
\end{equation}
We sample 5000 points around the two modes to estimate the posterior mass ratio. For each mode, the approximate posterior mass is
\begin{equation}
Z_k=\int_{mode~k} {\cal L}(\Theta)d\Theta\approx \frac{1}{N}\sum_i^N\frac{{\cal L}(\Theta_i)}{p(\Theta_i)},
\end{equation}
where $N$ is the number of sample points. For the proposal $p(\Theta)$, we use a Gaussian function to sample around the mode $\Theta_k$. The mean value is $\Theta_k$ and the standard deviation is $1\%$ of the mean value. However, for the parameters $T_1$ and $T_2$, we choose $0.1\%$ of the mean value because the likelihood function is very sensitive to changes in $T_1$ and $T_1$.  According to our calculation, the ratio of the posterior mass is $Z_2/Z_1\approx 0.1$. This means that the primary mode is statistically preferred over the secondary mode. Although our total number of photons is similar to that used in the real observation, we assumed a negligible background. In real observations, the background contributes a significant fraction of the detected photons. Thus, if the background is subtracted or fitted, the total number of photons of the effective data should be much smaller, which leads to a smaller data-to-noise ratio. From Fig.~\ref{fig:pulse-profiles} we can see that both modes can fit the data well and the parameter values are physical. In addition, this ratio changes with the spin frequency and the distance between the two modes in the parameter space. For the parameter inference of real observation with background models, systematic errors can exist in both background and pulse profile models, and it is more difficult to tell which mode is physically more inclined.

Finally, we try to identify possible secondary peaks for NICER observations based on our calculation. Priors on the observer inclination angle are included for the parameter inference of J0437-4715, J1231-1411 and J0614-3329. Thus, we focus on the fitting results for J0030+0451 and J0740+6620 pulse profiles~\citet{Miller2019,Riley2019,Miller2021,Riley2021}. For J0030+0451, we adopt the best-fit values of $u$, $\theta$, $\zeta_1$ and $\zeta_2$ given by parameter inference with ST-U model from~\citet{Miller2019,Riley2019}. However, the other solution for $u$ is too small to be considered as a reasonable value. For J0740+6620, we consider the fit result based on both NICER and XMM-Newton data. According to~\citet{Miller2021}, the best-fit values are $u=0.454$, $\theta=88.01^\circ$, $\zeta_1=137.45^\circ$ and $\zeta_2=107.49^\circ$. According to our calculation, the other peak can be found around $u=0.446$, $\theta=89.21^\circ$, $\zeta_1=138.26^\circ$ and $\zeta_2=69.88^\circ$. Unfortunately, only the value of $\zeta_2$ differs significantly from the original best-fit value. But this is still not distinguishable given the large uncertainty in $\zeta_2$ that can be seen in the corner plots of~\citet{Miller2021, Riley2021}. Maybe this explains why the uncertainty of $\zeta_2$ is so large. Thus, we also run another simulation in the appendix with parameters similar to this. However, the degeneracy we found does not affect the previous radius measurement by NICER.

\section{Summary}
\label{sec:summary}
Although it has been several years after the NICER observation. Reanalysis of NICER data and synthetic data can still get interesting results~\citep{Dittmann2024,Vinciguerra2023,Vinciguerra2024,Salmi2024a,Qi2025,Holt2025}. In this work, we extend our semi-analytic pulse profile model with two antipodal hot spots to the more general configuration with two non-antipodal spots; this generalization makes the model resemble the numerical ST-U models commonly used in analyses of observations from missions such as NICER and the upcoming eXTP. We find that our simplified semi-analytic model possesses an exact intrinsic degeneracy. This degeneracy may provide intuition for multi-modal posterior distributions discovered in previous works. In particular, a qualitative implication for future missions such as eXTP is that neutron stars with more informative priors are of greater value. For neutron stars in binary systems, inclination constraints from orbital observations can serve as priors to break these kinds of geometric degeneracies.

From the analytic expressions for the observed flux, we reveal an intrinsic degeneracy: two distinct sets of geometric parameter values (compactness, observer inclination angle, and colatitude of the two spots) can produce identical fluxes if the Doppler effect is neglected. We generate synthetic data to show that this degeneracy can still lead to two high likelihood regions corresponding to two modes on the likelihood surface even if Doppler effects and instrument response are considered. Statistically, the primary mode is more important than the secondary one for a 300~Hz neutron star according to our estimation of the posterior mass. The result also changes with the spin frequency, the number of data points, the noise level, and the distance between the two modes in the parameter space. For real observations with background models, whether this degeneracy identified here translates into a significant practical challenge for real observation depends on the extent to which it is broken or weakened by additional physics (finite spot size, atmospheres, instrumental response, etc.), and needs to be examined with full forward-modeling pipelines that include background treatments. 

\textbf{It is also interesting to consider the oblateness induced by rotation. For the spin frequency $\lesssim200$ Hz, the oblate Schwarzschild (OS) approximation is consistent with the S+D approximation and the exact numerical result~\citep{NICER2}. However, previous studies also showed that oblateness can lead to large deviations (the geometric oblateness effect can even rival the Doppler effect) if the spin frequency is between 300 and 500 Hz~\citep{Morsink2007,Cadeau2007}. In the thermal X-ray pulse profile modeling, where two hot spots and the real atmospheric beaming are considered, whether this effect is still important and how much it will break the degeneracy we found is beyond the scope of our paper with the S+D approximation that is widely used in NICER observations, and further study is needed.}

\section{Acknowledgments}
This research is supported by China's Space Origins Exploration Program. The authors thank Prof. Kejia Lee for important suggestions.

\appendix
\section{Posterior distributions with Increased Frequency and Data quality}\label{AppendixA}
\begin{figure*}[t]
	\centering
	\includegraphics[width=1\linewidth, keepaspectratio]{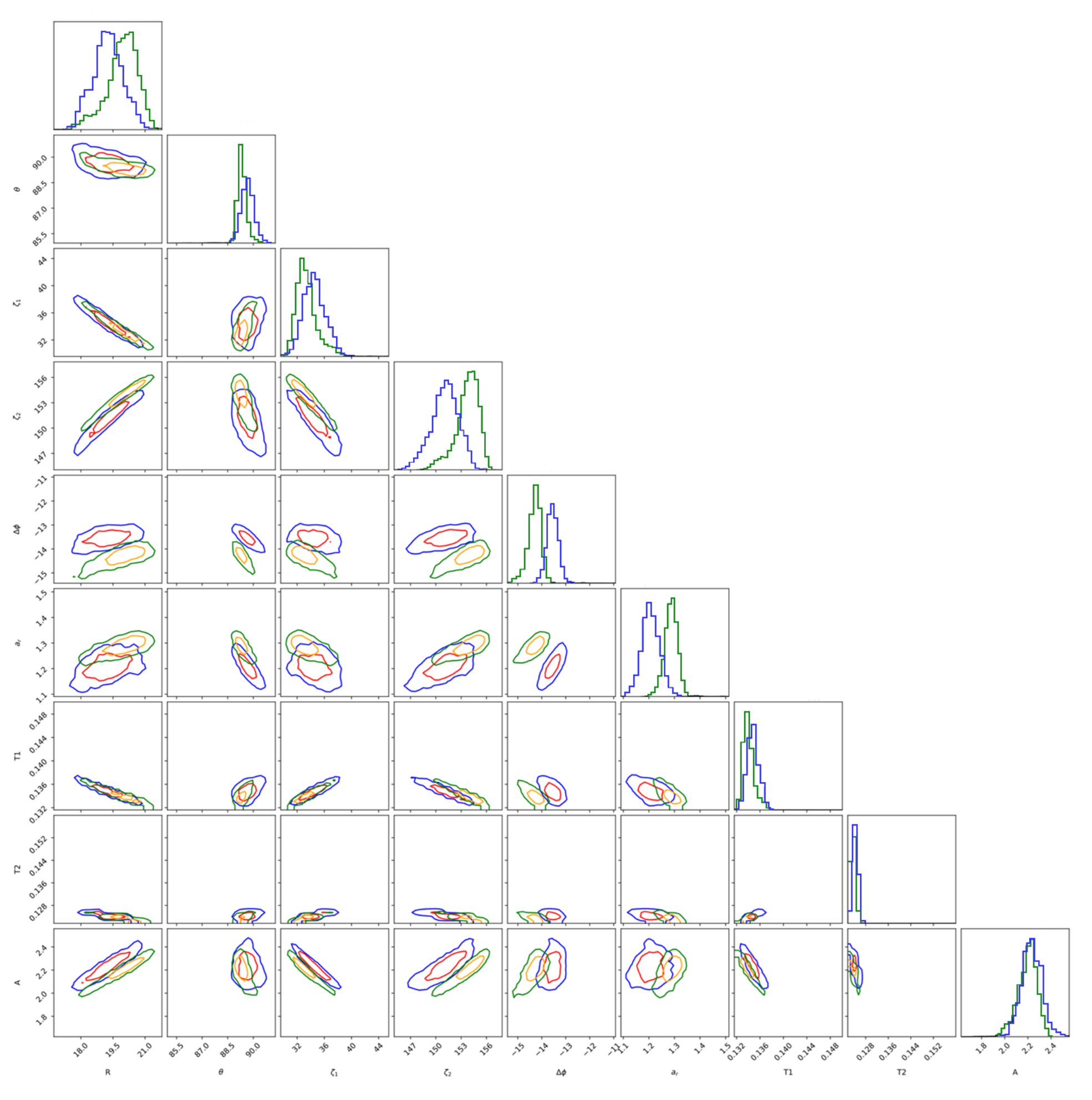}
	\caption{Contours of 68-th and 95-th percentiles of the posterior distribution around the secondary mode for parameter fits to pulse profiles at different frequencies. Red and blue contours and the blue histograms are for $f=200$~Hz, orange and green contours and the green histograms are for $f=400$~Hz.
    }
    \label{fig:frequecy}
\end{figure*}

In this Appendix, we generate more synthetic pulse profiles with different parameters to show how the second peak on the likelihood surface varies with frequency and total number of photons. We generate synthetic pulse profiles at frequencies of 200~Hz and 400~Hz, as well as pulse profiles at 300~Hz, but with doubled total number of photons (achieved by increasing the parameter $A$), and then perform the same fitting procedure. The other parameter values are the same as those used in Section~\ref{sec:multimodal}. As the frequency increases, the secondary peak on the posterior surface does not disappear because there is still a local likelihood maximum. However, the position of the mode and the posterior mass ratio of the primary mode to the secondary mode change. For $f=200$~Hz, the approximate posterior mass ratio of the secondary mode to the primary mode becomes 0.5. For $f=400$~Hz, this value decreases to 0.08. The contours of the 68-th and 95-th percentiles around the secondary mode are shown in Fig.~\ref{fig:frequecy}. We can see that the position of the secondary mode also changes with the spin frequency. If we double the total number of photons, the approximate posterior mass ratio decreases significantly to 0.01. Therefore, the statistical error of the data plays an important role in whether we can distinguish the two modes statistically. Thus, whether this degeneracy can result in multi-modal structures in real observation strongly depends on the quality of data and the background level.

\begin{figure*}[t]
	\centering
	\includegraphics[width=1\linewidth, keepaspectratio]{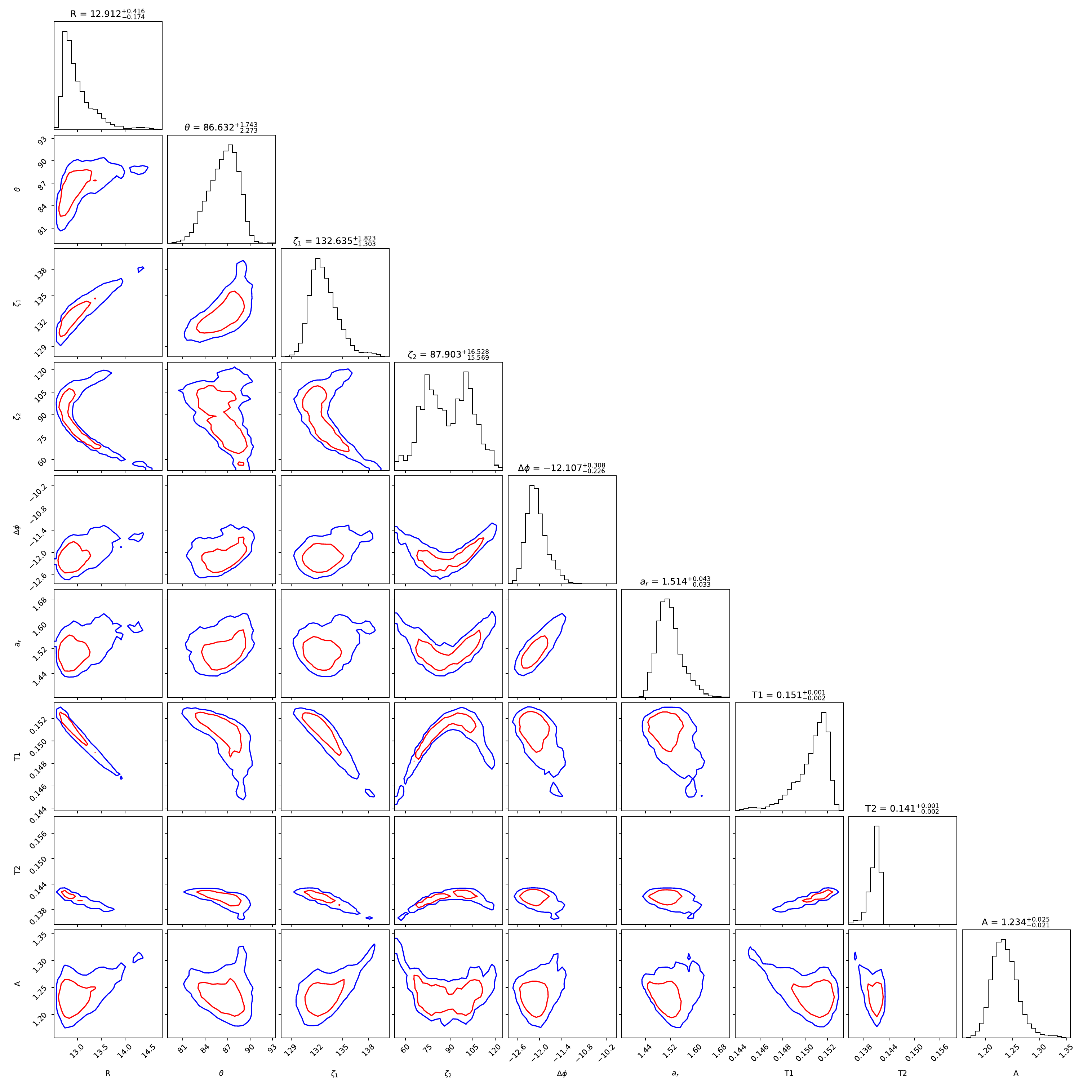}
	\caption{Contours of 68-th and 95-th percentiles of the posterior distribution for parameter fits to pulse profiles at $f=200$~Hz. The synthetic pulse profile is generated with a special set of parameters: $u=0.45$, $\theta=88^\circ$, $\zeta_1=135^\circ$ and $\zeta_2=105^\circ$. According to our calculation, another high-likelihood mode is close to this ground truth with similar $u$, $\theta$, $\zeta_1$ but much smaller $\zeta_2$. This leads to a large uncertainty in the measurement of $\zeta_2$.
    }
    \label{fig:closemodes}
\end{figure*}

In addition, we also want to show that when the two degenerate solutions are close to each other, the measurement result will have a large uncertainty that is not due to the statistic error. We generate synthetic pulse profiles with $u=0.45$, $\theta=88^\circ$, $\zeta_1=135^\circ$ and $\zeta_2=105^\circ$ and keep other parameters the same. Then, according to our calculation, the other high likelihood region should be around $u=0.442$, $\theta=89.13^\circ$, $\zeta_1=135.797^\circ$ and $\zeta_2=72.25^\circ$ with similar $u$, $\theta$, $\zeta_1$ but much smaller $\zeta_2$. This leads to a large uncertainty in the measurement of $\zeta_2$ in Fig.~\ref{fig:closemodes}.

\bibliographystyle{apj}
\bibliography{MS}

\end{document}